\documentclass[conference]{IEEEtran}
\usepackage{bm}
\usepackage{amsmath}
\usepackage{amsfonts}
\usepackage{cite}
\usepackage{graphicx}
\usepackage{subcaption}
\usepackage{nomencl}
\usepackage{color}
\usepackage{float}
\usepackage{soul, xcolor}
\usepackage{siunitx}
\DeclareMathOperator{\argmax}{\arg\max}

\newcommand{\ex}{\mathbb{E}}
\DeclareMathOperator{\heav}{H}

\ifCLASSINFOpdf
\else
\fi
\begin{document}
\setstcolor{red}

%
\title{A Principal-Agent Model of Systems Engineering Processes with Application to Satellite Design}


%
\author{\IEEEauthorblockN{Salar Safarkhani\IEEEauthorrefmark{2},
Vikranth Reddy Kattakuri\IEEEauthorrefmark{2},
Ilias Bilionis\IEEEauthorrefmark{2}\IEEEauthorrefmark{1}, Jitesh Panchal\IEEEauthorrefmark{2}}
\IEEEauthorblockA{\IEEEauthorrefmark{2}School of Mechanical Engineering, Purdue University, West Lafayette, Indiana 47907-2088}
\IEEEauthorblockA{\IEEEauthorrefmark{1}Corresponding author, Email: ibilion@purdue.edu}
}


\maketitle

\begin{abstract}
We present a principal-agent model of a one-shot, shallow, systems engineering process. The process is ``one-shot'' in the sense that decisions are made during one time step and that they are final.
The term ``shallow'' refers to a one-layer hierarchy of the process.
Specifically, we assume that the systems engineer has already decomposed the problem in subsystems, and that each subsystem is assigned to a different subsystem engineer.
Each subsystem engineer works independently to maximize their own expected payoff.
The goal of the systems engineer is to maximize the system-level payoff by incentivizing the subsystem engineers.
We restrict our attention to requirement-based system-level payoffs, i.e., the systems engineer makes a profit only if all the design requirements are met.
We illustrate the model using the design of an Earth-orbiting satellite system where the systems engineer determines the optimum incentive structures and requirements for two subsystems: the propulsion subsystem and the power subsystem. 
The model enables the analysis of a systems engineer's decisions about optimal passed-down requirements and incentives for sub-system engineers under different levels of task difficulty and associated costs.
Sample results, for the case of risk-neutral systems and subsystems engineers, show that it is not always in the best interest of the systems engineer to pass down the true requirements.
As expected, the model predicts that for small to moderate task uncertainties the optimal requirements are higher than the true ones, effectively eliminating the probability of failure for the systems engineer.
In contrast, the model predicts that for large task uncertainties the optimal requirements should be smaller than the true ones in order to lure the subsystem engineers into participation.
\end{abstract}


%
\IEEEpeerreviewmaketitle

\section{Introduction}
The ubiquitous problem of schedule and cost over-runs during 
the development of large-scale complex systems is well 
documented within systems engineering 
literature~\cite{collopy_value-driven_2011}. Various remedies 
have been proposed to address these unsustainable trends, including, better methods 
and tools for managing complexity, better incentive mechanisms, and transition 
from document-based systems engineering to model-based systems engineering 
(MBSE). The research community has related these trends to the fundamental way 
in which systems engineering processes are carried out. Requirements 
engineering, which is one of the foundational processes within systems 
engineering, has been identified as a key source of the inefficiency. Collopy, 
for example, argues that the use of requirements in systems engineering is an 
ineffective way of coordination between systems engineer and subsystems 
engineers~\cite{collopy_adverse_nodate}. Therefore, there is a need within 
systems engineering to model and analyze the requirements engineering process.

There have been few efforts in addressing this need. By modeling systems engineering processes as multi-disciplinary design optimization problems, Collopy et al. show that using requirements within systems engineering processes creates design trade conflicts among different subsystems, resulting in dead losses within the system~\cite{collopy_adverse_nodate}.
Collopy and Hollingsworth~\cite{collopy_value-driven_2011} espouse the use of value driven design (VDD) as a better alternative to requirements engineering, wherein, objectives for extensive attributes are passed down instead of requirements. Their model is applicable for settings where the incentives of the subsystem engineers are well aligned with the objectives of the systems engineers. This assumption may be valid when both the systems engineers and sub-systems engineers are within the same organization. If, on the other hand, the subsystems engineers are independent decision makers with private information and driven by their own objectives, their model is inappropriate, and superiority of VDD is not clear. 

To model realistic systems engineering processes, there is a need to model interactive decisions of self-interested actors using game theory. Vermillion and Malak~\cite{vermillion_using_2015} take initial steps in that direction by modeling the interactions between a systems engineer and subsystems engineers using principal-agent models. They adapt the generalized principal-agent model to model the situation of a systems engineer delegating work to subsystem engineers as a one-shot game.
Their adaption is primarily focused on incorporating behavioral aspects such as deviations from expected utility maximization within the general principal-agent model.

While incorporating behavioral aspects within principal-agent models is an important step forward, we believe that there is still a lack of models that account for unique aspects of systems engineering, specifically, the information available to systems engineers, the state of technology, the uncertainty in the ability to achieve specific outcomes, and the level of difficulty of the tasks. 
To address this gap, we develop a principal-agent model~\cite{meluso_gaming_2017} of a simple systems engineering process in 
which decisions are made once and all involved individuals have their own 
private interests. 
Note that our model is an oversimplification of real systems engineering processes which are iterative and in which information and outcomes flow back and forth between the systems engineer and subsystem engineers until a final decision is made.
Our model should be considered as a first step towards modeling full fledged systems engineering processes.
Using our framework, we study the optimal mechanisms within the class of requirement-based incentives.
We illustrate the model using a satellite design case study with two subsystems: power and propulsion.
Specifically, we show how historical data can be used to infer the parameters of our process model to this case study.

The paper is organized as follows.
We start Sec.~II by describing our systems engineering process model in general terms and we cast the selection of subsystem incentives as a mechanism design problem.
In Sec.~II.A, we justify some assumptions for the nature of the subsystem engineers which simplify the model.
In Sec.~II.B, we focus the discussion on the class of requirement-based incentives for all agents.
In Sec.~II.C and~II.D, we non-dimensionalize the equations and we describe how the parameters can be inferred from readily available historical data.
In Sec.~III, we apply the model to the design of a spacecraft taking into account two subsystems (power and propulsion).
Finally, in Sec.~IV, we present our numerical study and in Sec.~V our conclusions.

\section{Modeling a Single-Shot, Shallow Systems Engineering Process}
We consider a model of a \emph{single-shot} (evolves in one time-step), \emph{shallow} (considers one-layer interactions between a systems engineer and multiple subsystem engineers) systems engineering process.
The systems engineer (SE) has already decomposed the problem in $N$ subsystems.
The SE assigns the design of each subsystem to a subsystem engineer (sSE).
Each sSE designs independently to maximize their own expected payoff, and returns the design outcome back to the SE.
The goal of the SE is to incentivize the sSEs to produce subsystem designs that maximize the expected system-level payoff by choosing appropriate contracts.
We start by formulating the problem of optimal subsystem contract design in its full generality.
Then, we make simplifying assumptions about the form of the sSEs' quality and utility functions, and, we study the optimality requirement-based contracts.

Let $i=1,\dots,N$ be a label indexing the sSEs.
The $i$-th sSE chooses a normalized \emph{effort level} $e_i\in[0,1]$.
This measures the percentage of maximum effort that the sSE can allocate to this specific project within a predefined time framework, e.g., in a fiscal year.
Second, the units of the effort depend on the nature of the sSE.
If the sSE is an individual that works for the same organization as the SE, then the effort $e_i$  could be measured in terms of the percentage of that the individual dedicates to the project.
Alternatively, if the sSE is an external contractor, e.g., another company, then effort could be measured in terms of the percentage of the available yearly resources that the contractor dedicates to this particular project.
We denote the \emph{cost of effort} to the sSE as $c_i(e_i)$.
In economic terms, $c_i(e_i)$ is an \emph{opportunity cost}, i.e., the monetary gain the sSE could receive, but forfeits, to participate in this particular project.

Let $(\Omega, \mathcal{F}, \mathbb{P})$ be a probability space associated with random states of nature.
We model the design qualities that the $i$-th sSE can produce as a stochastic process $q_i(e_i,\omega)$.
That is, the \emph{quality function} $q_i(e_i,\omega)$ gives the design quality that the $i$-th sSE can achieve by choosing an effort level $e_i$ if the state of nature is $\omega\in\Omega$.
The quality function is normalized so that zero corresponds to the quality of the current state-of-the-art.
Note that we have deliberately chosen to ignore the dependence of $q_i(e_i, \omega)$ on any private information.
In other words, we assume that the form of the quality function is common knowledge.

Each of the sSEs enters a contract with the SE.
The contracts describe \emph{transfer functions}, $t_i(q_i)$, which specify the transport of monetary funds from the SE to the sSE contingent on the quality of the design that the sSE produces.
Therefore, the \emph{payoff} to the sSE is:
\begin{equation}
\label{pisub}
\pi_i\left(e_i,\omega\right) = t_i\left(q_i\left(e_i,\omega\right)\right)-c_i\left(e_i\right).
\end{equation}
We assume that the sSE selects an optimal effort level ex-ante, i.e., before they observe the future state of nature $\omega$.
If we further assume that the sSE is risk-neutral, then rationality implies that they should select their effort level by maximizing their expected payoff:
\begin{equation}
\label{exsub}
{e_i}^*\left(t_i\left(\cdot\right)\right) = \underset{e_i\in[0,1]}\argmax\ \ex_{\omega}\left[\pi_i\left(e_i,\omega\right)\right].
\end{equation}
Note the dependence on the transfer function.

To ensure that the sSEs are willing to participate in this project, their optimal expected payoff must be positive.
Otherwise, the sSEs have no incentive to be part of the project as their expected monetary benefit is smaller than their opportunity cost.
Therefore, the SE must choose transfer functions that enforce the \emph{participation constraints}:
\begin{equation}
\label{participationconstraints}
\mathbb{E}_\omega\left[\pi_i\left({e_i}^*\left(t_i\left(\cdot\right)\right),\omega\right)\right] \ge 0,
\end{equation}
for $i=1,\dots,N$.

The SE obtains from each sSE the following design qualities:
\begin{equation}
\label{optimalquality}
{q_i}^*\left(t_i\left(\cdot\right),\omega\right)=q_i\left({e_i}^*\left(t_i\left(\cdot\right)\right),\omega\right).
\end{equation}
If $V(q_1,\dots,q_N)$ is the net present value of any cash flows that result from a system with subsystem qualities $q_1,\dots,q_N$, then the total payoff to the SE is:
\begin{equation}
\label{pisys}
\begin{aligned}
\pi\left(\mathbf{t}\left(\cdot\right),\omega\right)&=V\left({q_1}^*\left(t_1\left(\cdot\right),\omega\right),\cdots,{q_N}^*\left(t_N\left(\cdot\right),\omega\right)\right) \\
&-\sum_{i=1}^{N}t_i\left({q_i}^*\left(t_i\left(\cdot\right),\omega\right)\right),
\end{aligned}
\end{equation}
where we defined $\mathbf{t\left(\cdot\right)}=\left(t_1\left(\cdot\right),\cdots,t_N\left(\cdot\right)\right)$.
Assuming that the SE makes ex-ante, risk-neutral decisions, they must select transfer functions that solve:
\begin{equation}
\label{sysmax}
\mathbf{t}^*\left(\cdot\right)=\underset{\bm{t\left(\cdot\right)}}\argmax\ \ex_{\omega}\left[\pi\left(\mathbf{t}\left(\cdot\right),\omega\right)\right],
\end{equation}
subject to the $N$ participation constraints defined in Eq.~(\ref{participationconstraints}).
Of course, if the optimal expected payoff $\mathbb{E}_\omega[\pi(\mathbf{t}^*(\cdot),\omega)]$ is negative, then the SE does not initiate the project in the first place.
In what follows, we study this \emph{mechanism design} problem by making specific assumptions for the form of the SE value function, the sSEs' quality functions and costs, and the form of the possible transfer functions.

\subsection{Assumptions about the subsystem engineers}
The random field $q_i(e_i, \omega)$ captures the common state of knowledge about what is technologically possible in the design quality of subsystem $i$.
Using the Karhunen-Lo\`eve expansion \cite{ghanem_stochastic_1991}, the random field $q_i(e_i,\omega)$ can be written as:
\begin{equation}
q_i(e_i,\omega) = q_i^0(e_i) + \sum_{k=1}^\infty \sqrt{\lambda_{ik}}\xi_{ik}(\omega)\phi_{ik}(e_i),
\end{equation}
where $q_i^0(e_i)$ is the mean of the random field, $\lambda_{ik}$, $\phi_{ik}(e_i)$ are the eigenvalues and eigenvectors of its covariance function, respectively, and the random variables $\xi_{ik}$ are zero mean, unit variance, and uncorrelated.
Assuming stationarity of the process, these quantities can, in principle, be estimated statistically from historical data of marginal investments versus increases in product quality.
As a first approximation, we truncate the series at $k=1$ keeping only the largest eigenvalue:
$$
q_i(e_i,\omega) \approx q_i^0(e_i) + \sqrt{\lambda_{i1}}\xi_{i1}(\omega)\phi_{i1}(e_i).
$$
Furthermore, we approximate the zero-mean and unit-variance random variable $\xi_{i1}$ as a standard normal random variable $\xi$ (the standard normal is the maximum entropy distribution with zero-mean and unit variance).
We also assume that the first eigenvector is approximately constant, $\phi_{i1}(e_i) \approx \mbox{const}$, and we introduce the new variable $\sigma_i = \sqrt{\lambda_i}\phi_{i1}(e_i)$.
Without loss of generality, we may take that $\sigma_i > 0$.
Finally, we take the first order Taylor expansion of $q_i^0(e_i) = a_i e_i + O(e_i^2)$, recalling that we scale the quality so that zero corresponds to the current state-of-the-art, which can be delivered without any effort.
This is reasonable since we are considering a one-shot systems engineering process which, necessarily, takes place in a limited amount of time.
For larger timescales, we expect $q_i(e_i,\omega)$ to be curved: Concave for mature technologies, and convex followed by concave for emerging technologies.
Furthermore, we take $a_i > 0$ since more effort can only lead to increased design quality.
To summarize, we model the quality as:
\begin{equation}
\label{qsub}
q_i\left(e_i,\omega\right)=a_i e_i + \sigma_i \xi\left(\omega\right),
\end{equation}
where $a_i,\sigma_i>0$, and $\xi \sim N\left(0,1\right)$.

The parameter $a_i$ depends on the skills of the sSE as well as on the maturity of the underlying technology.
That is, a skillful sSE produces a higher increase in quality from the state-of-the-art than a less skilled sSE.
Therefore, keeping the maturity of the technology fixed, $a_i$ expected to grow as the skills of the sSE are improved.
Similarly, keeping the skills of the sSE fixed, $a_i$ decreases as a function of the maturity of the underlying technology.
The more mature the underlying technology is, the more difficult it becomes to obtain a given increase in quality.

The parameter $\sigma_i$ behaves in exactly the opposite way.
A skillful sSE produces design qualities that vary less, therefore $\sigma_i$ decreases as skill improves.
On the other hand, we expect that subsystem designs that depend on mature technologies are more predictable, therefore $\sigma_i$ decreases as technological maturity increases.

Our final assumption is that the sSE's cost grows linearly with effort as:
\begin{equation}
\label{csub}
c_i\left(e_i\right)=c_i e_i,
\end{equation}
where $c_i > 0$.
This assumption is reasonable for almost all types of sSEs.
In particular, when sSE is an individual engineer, $c_i$ could be the average industry salary per unit effort.
For a contractor, $c_i$ could be the expected payoff per unit effort of the next most profitable project that they could be engaging in.

\subsection{Optimal requirement-based incentives}
We assume that the SE has $N$ requirements, $r_1,\dots,r_N$, one to be satisfied by each subsystem.
These requirements arise from the \emph{business} case of the project.
Mathematically, the system design is successful if $q_i > r_i$ for all $i=1,\dots,N$.
If the value of a successful system design is $V_0$, then the value function of the SE is:
\begin{equation}
\label{reqval}
V(q_1,\dots,q_N) = V_0 \prod_{i=1}^N \heav\left(q_i-r_i\right),
\end{equation}
where $\heav\left(\cdot\right)$ is the Heaviside function:
$$
\heav(x) = \begin{cases}
1,\;\text{if}\;x\ge 0,\\
0,\;\text{otherwise}.
\end{cases}
$$

We restrict our attention to the study of \emph{requirement-based} transfer functions:
\begin{equation}
t_i\left(q_i;\boldsymbol{\psi}_i\right) = \psi_{i1}+\psi_{i2}\heav\left(q_i-\psi_{i3}\right).
\end{equation}
The first parameter, $\psi_{i1}$, specifies the amount that is going to be paid simply for agreeing to participate in the project, guaranteeing a design quality at least the same as the current state-of-the-art.
The second parameter, $\psi_{i2}$, specifies the amount to be paid if the subsystem engineer meets specific requirements.
The third parameter, $\psi_{i3}$, specifies the requirement that the subsystem engineer has to meet.
Note that the optimal \emph{passed-down} requirement is, in general, different than the real requirement, $r_i$.

We start with the optimal decision of the $i$-th sSE given a fixed contract $t_i\left(q_i;\boldsymbol{\psi}_i\right)$.
The payoff is:
$$
\pi_i(e_i,\omega) = \psi_{i1}+\psi_{i2}\heav\left(a_ie_i + \sigma_i\xi_i(\omega)-\psi_{i3}\right) - c_ie_i.
$$
To take the expectation over $\omega$, we use the following result:
\begin{equation}
\label{phiformula}
\ex_{\omega}\left[\heav\left(\lambda+\sigma\xi(\omega)\right)\right]
= \Phi\left(\frac{\lambda}{\sigma}\right),
\end{equation}
where $\Phi(\cdot)$ is the cumulative distribution function (CDF) of a standard normal random variable.
Taking the expectation over $\omega$, we get:
$$
\ex_\omega\left[\pi_i(e_i,\omega)\right]=\psi_{i1}-c_ie_i+\psi_{i2}\Phi\left(\frac{a_ie_i-\psi_{i3}}{\sigma_i}\right).
$$
So, the optimal effort level is:
\begin{equation}
e_i^*\left(t_i\left(\cdot;\boldsymbol{\psi}_i\right)\right)= \underset{{e_i\in[0,1]}}\argmax \left\{\psi_{i1}-c_ie_i+\psi_{i2}\Phi\left(\frac{a_ie_i-\psi_{i3}}{\sigma_i}\right)\right\}.
\end{equation}
We solve this problem using Brent's method~\cite{brent_algorithms_2013} as implemented in SciPy~\cite{jones_scipy:_2001}.

The SE problem consists of maximizing:
\begin{equation}
\label{seexppi}
\begin{aligned}
\ex_\omega\left[\pi\left(\mathbf{t}\left(\cdot,\boldsymbol{\psi}\right),\omega\right)\right] =& V_0\prod_{i=1}^N\Phi\left(\frac{a_ie_i^*\left(t_i\left(\cdot;\boldsymbol{\psi}_i\right)\right) - r_i}{\sigma_i}\right)\\
&-\sum_{i=1}^N\psi_{i1}\\
&-\sum_{i=1}^N\psi_{i2}\Phi\left(\frac{a_ie_i^*\left(t_i\left(\cdot;\boldsymbol{\psi}_i\right)\right) - \psi_{i3}}{\sigma_i}\right),
\end{aligned}
\end{equation}
subject to the participation constraints:
\begin{equation}
\psi_{i1}+\psi_{i2}\Phi\left(\frac{a_ie_i^*\left(t_i\left(\cdot;\boldsymbol{\psi}_i\right)\right) - \psi_{i3}}{\sigma_i}\right)-c_ie_i^*\left(t_i\left(\cdot;\boldsymbol{\psi}_i\right)\right)\ge 0,
\end{equation}
and the bounds:
\begin{equation}
\psi_{ik}\ge 0.
\end{equation}
for $k=1,2,3$ and $i=1,\dots,N$.
In practice, the optimal solution is always within the following bounds:
\begin{equation}
0\le \psi_{i1},\psi_{i2} \le 2c_i,\;\text{and}\;r_i-3\sigma_i<\psi_{i3}<r_i+3\sigma_i.
\end{equation}
We solve this problem using the constrained optimization by linear approximation (COBYLA) algorithm~\cite{lucidi_global_2002} as implemented in~\cite{jones_scipy:_2001}. 

\subsection{Non-dimensionalization of the equations}
Without loss of generality, we can pick all the subsystem requirements to be $r_i=1$.
This can be achieved by appropriately scaling all $a_i$'s and all $\sigma_i$'s.
In other words, the quality function of each subsystem will be measured in units of the corresponding business-imposed requirement.
Then, the inverse of the coefficient $a_i$, can be interpreted as the effort that needed to meet the subsystem requirement for sSE if there is no uncertainty.
We will consider two levels of $a_i$'s corresponding to different levels of subsystem design difficulty: (i) (hard) $a_i=1.5$, (ii) (easy) $a_i=2$.
Having scaled the quality function in this way, the variance parameter $\sigma_i$ can also be interpreted as the amount of uncertainty in the quality of the final design as a percentage of the requirement.
We will consider three levels of uncertainty: (i) (low) $\sigma_i=0.05$, (ii) (moderate) $\sigma_i=0.1$, and (iii) (high) $\sigma_i=0.2$.

Finally, also without loss of generality, we may set $V_0=1$.
This can be achieved by appropriately scaling the transfer functions and the opportunity costs of all sSEs.
It amounts to measuring all monetary quantities, in terms of the SE's maximum value.
For the opportunity costs of the sSEs, we will consider two levels: (i) (low) $c_i=0.01$, (ii) (high) $c_i=0.05$.

\subsection{Extracting subsystem parameters from historical data}
To study decision making in the context of a real application, one needs to extract all parameters, $a_i,c_i,\sigma_i,\dots$, from historical data.
To this end, let $Q_i$ denote the quality of the $i$-th subsystem in physical units, and $I_i$ the cumulative investment per firm on this technology.
Historical data, say $\mathcal{D}_i = \left\{\left(I_{is}, Q_{is}\right)\right\}_{s=1}^{S_i}$, of these quantities are readily available for many technologies.
We model the relationship between $Q_i$ and $I_i$ as:
\begin{equation}
\label{Q_of_I}
Q_{i} = Q_{i0} + A_{i}(I_{i}-I_{i0})+ \Sigma_{i}\xi_{i}(\omega), 
\end{equation}
where  $Q_{i0}$ and $I_{i0}$ is the current state of these variables, $\xi_i(\omega)\sim\mathcal{N}(0,1)$, and $A_i$ and $\Sigma_i$ are parameters to be estimated from the all available data $\mathcal{D}_i$.
We use a maximum likelihood estimator~\cite{tipping_probabilistic_1999} for $A_i$ and $\Sigma_i$.
This is equivalent to least squares estimate for $A_i$:
\begin{equation}
\hat{A}_i = \arg\min_{A_i}\sum_{s=1}^{S_i}\left[Q_{i0}+A_i(I_{is}-I_{i0}) - Q_{is}\right]^2,
\end{equation}
and to setting $\Sigma_i$ equal to the mean residual square error:
\begin{equation}
\hat{\Sigma}_i = \frac{1}{S_i}\sum_{s=1}^{S_i}\left[Q_{i0}+A_i^*(I_{is}-I_{i0}) - Q_{is}\right]^2.
\end{equation}

Now, let $Q_{ir}$ be the required quality for subsystem $i$ in physical units.
The scaled quality of a subsystem $q_i$, can be defined as:
\begin{equation}
\label{q_def}
q_{i} = \frac{Q_{i} - Q_{i0}}{Q_{ir} - Q_{i0}}, 
\end{equation}
with this definition, we get $q_i=0$ for the state-of-the-art, and $q_i=1$ for the requirement.
Substituting Eq.~(\ref{Q_of_I}) in Eq.~(\ref{q_def}) while making use of the maximum likelihood estimates for $A_i$ and $\Sigma_i$, we get
\begin{equation}
\label{q_intermediate}
q_{i} = \frac{\hat{A}_{i}}{Q_{ir} - Q_{i0}}(I_{i}-I_{i0})+ \frac{\hat{\Sigma}_{i}}{Q_{ir} - Q_{i0}}\xi_i(\omega).
\end{equation}
From this equation, we see that the uncertain parameter of our previous discussions, can be obtained from
\begin{equation}
\sigma_i = \frac{\hat{\Sigma}_{i}}{Q_{ir} - Q_{i0}}.
\end{equation}

Finally, let $T_i$ and $C_i$ represent the time for which the $i^{th}$ sSE is to be hired and the cost of the engineer per unit time, respectively.
$T_i$ is just the duration of the systems engineering process we consider.
The opportunity cost $C_i$ can be read inferred from the balance sheets of publicly traded firms related to the technology.
We can now define the effort variable $e_i$ as:
\begin{equation}
\label{e_def}
e_{i} = \frac{I_{i}-I_{i0}}{T_{i}C_{i}}.
\end{equation}
From this and Eq.~(\ref{q_intermediate}), we see that $a_i$ is given by:
\begin{equation}
\label{ai_value}
a_i = \frac{T_iC_i\hat{A}_i}{Q_{ir}-Q_{i0}}.
\end{equation}

\section{Illustrative Example: Spacecraft Design Problem}
\subsection{Spacecraft systems design}
During the initial proposal phase of satellite development for scientific applications, the principal investigator puts forward an estimate of how the goals of the project will be achieved through engineering means.
These project goals concern the overarching science objectives of the mission which correspond to the instrument design at the heart the satellite. 
However, in the process of successfully launching a scientific instrument into space, all the necessary components to power the instruments, actuate the spacecraft, and transmit data must also be included in the satellite payload.
These mission requirements are translated to specific functional requirements for each subsystem of the spacecraft.

Typically, a spacecraft consists of seven main sub-systems~\cite{wertz_space_2011}, namely, electrical power subsystem (EPS), propulsion, attitude determination and control (ADC), on-board processing, telemetry, tracking and command (TT\&C), structures and thermal subsystems. 
For the proposed study, we will focus our attention on two subsystems ($N=2$): EPS and propulsion.
Simplifying the analysis, we assume that the design of these subsystems will 
be assigned to two (subsystem) engineers in a one-shot fashion. Note that, the systems engineering process of the spacecraft 
is an iterative process and the information and results are exchanged and flow 
back and forth between the SE and sSEs in each iteration.
Our model is a very crude approximation of reality. The goal of 
the SE is to optimally incentivize the sSEs to produce subsystem designs that 
meet the mission's requirements.



\subsection{Electrical power subsystem}
The EPS is designed and configured to perform several key functions, the primary being a continuous and reliable source of peak and average
electrical power for the life of the mission.
It consists of a power source, energy storage, power conversion/distribution and power regulations and control equipment.
Typically, for earth orbiting satellites, one employs solar photo-voltaic arrays as a primary energy source and batteries as secondary power storage units. 
Silicon solar cells are the most commonly used photo-voltaic cells for space applications because of their low cost and high availability.

In this study, the design quality of interest for the EPS sSE is chosen to be the solar cell efficiency, i.e., $Q_1$ is the efficiency of Si-based solar photo-voltaic cells expressed as a percentage value, and $I_1$ is the average cumulative investment in solar cell research per firm.
To estimate the relationship between $Q_1$ and $I_1$, we first considered the global trends of commercial Si module efficiencies and cumulative investments in solar-cell research over 2001--2008.
Over this period, a total of hundred solar cell research and development companies received venture capital (VC) funding~\cite{noauthor_historical_nodate}.
Based on this information, we assume that, on average, thirteen companies are participating in solar cell R\&D every year. 

\begin{figure}[h!]
\centering
\includegraphics[scale=1.0]{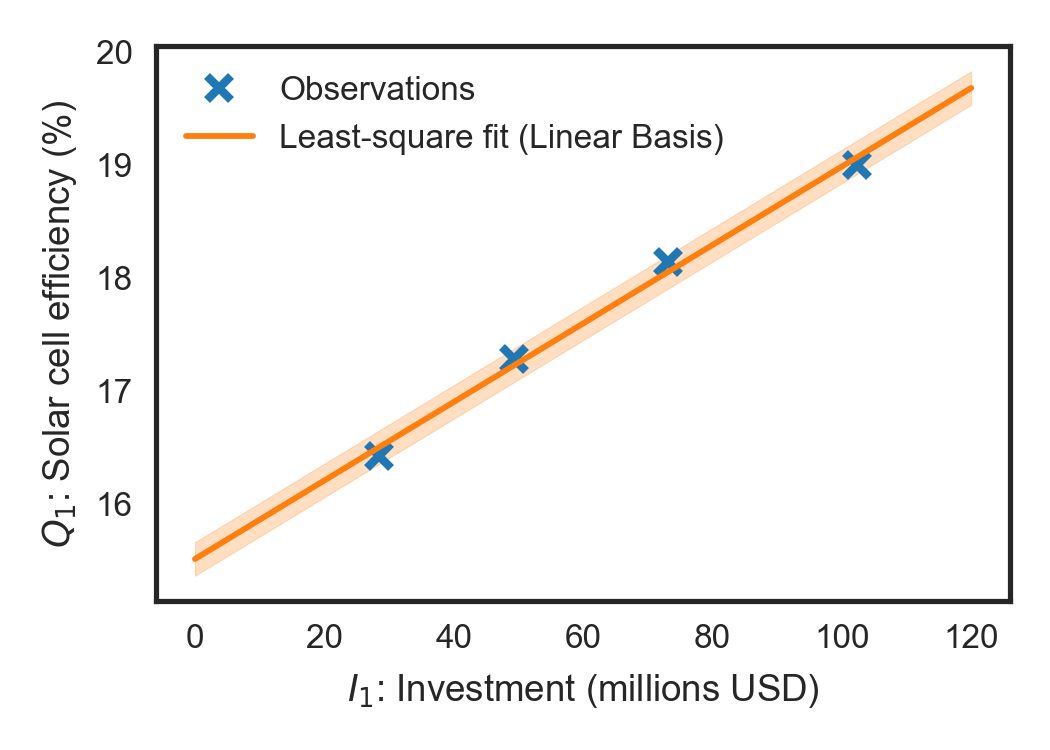}
 \caption{
 Spacecraft case study (EPS subsystem): Historical data (2001--2008) of solar cell efficiencies vs cumulative investment per firm.
The solid line and the shaded area correspond to the maximum likelihood fit of a linear regression model and the corresponding $95\%$ prediction intervals, respectively. 
}
\label{epsdata3}
\end{figure}

The variation of commercial crystalline-Si module efficiency ($Q_1$ (\%) ) with time is taken from~\cite{osti_983330}.
The trend of cumulative global VC investment per firm ($I_1$ (millions USD)) in crystalline Si-cell technology over time is obtained from~\cite{noauthor_historical_nodate}.
In 2008, the state-of-the-art was $Q_{10}=19\%$ and the cumulative investment per firm was $I_{10} = 102.4$ million USD.
A maximum likelihood fit of the parameters in Eq.~(\ref{Q_of_I}) results in a regression coefficient of $A_1=0.035\%$ per million USD, and standard deviation $\Sigma_1 = 0.15\%$.
We can visualize the data and the maximum likelihood fit in Fig.~\ref{epsdata3}.

The cost parameter $C_1$ is estimated by considering the total pay towards employees salaries over the number of employees in R\&D jobs in global PV industry.
According to the statistics~\cite{osti_983330}, around 2,320 employees were involved in such jobs by the end of 2008.
The value of $C_1$ is the median salary of a solar cell development engineer which is approximately equal to 100,000 (0.1 million) USD based on data from~\cite{eps_median_sal}. 
Substituting the values of $Q_{10}$, $C_1$ and $A_1$ in Eq.~\ref{ai_value} yields the following relationship between the parameters $Q_{1r}$, $T_1$ and $a_1$:
\begin{equation}
\label{a1}
a_{1} = \frac{0.0035 T_{1}}{Q_{1r}-19}  
\end{equation}

\begin{figure}[h!]
\centering
\includegraphics[scale=1.0]{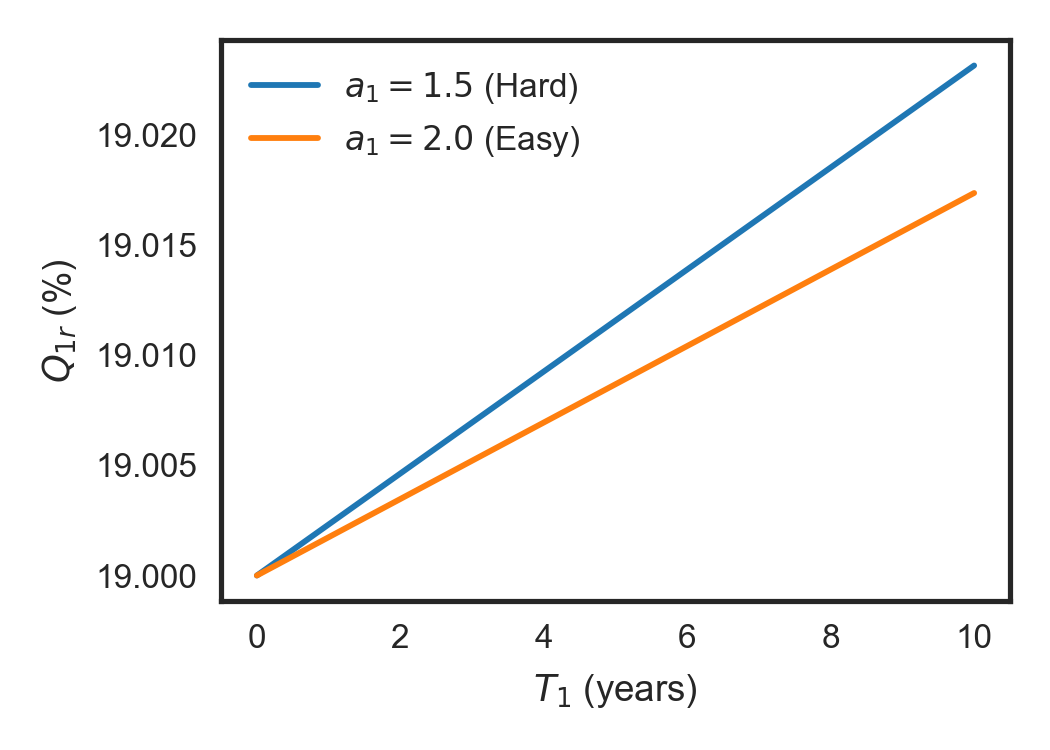}
 \caption{Spacecraft case study (EPS subsystem): Variation of $Q_{1r}$ (percentage efficiency) w.r.t $T_1$ (time in years)}
\label{epscontour}
\end{figure}

The variation of required quality (efficiency) for sSE-1 ($Q_{1r}$~(\%)) with respect to the time for which one person from sSE firm is to be hired ($T_1$ years) is shown in Fig.~\ref{epscontour}, for two different levels of EPS design difficulty.

\subsection{Propulsion subsystem}
Space propulsion systems essentially provide thrust to lift the launch vehicle along with its payload from the launch pad and place the payload into low-Earth orbit. They assist payload transfer between lower and higher orbits or into trajectories based on the type of mission. Finally, they provide thrust for attitude control and orbit corrections~\cite{wertz_space_2011}. Based on mission profiles, performance requirements for propulsion systems include thrust, total impulse, and duty cycle specifications for which, the specific impulse and propellant density are the key parameters. 

Chemical combustion systems, which are the most common systems for space applications can be divided into three basic categories: liquid. solid and hybrid. The terminology refers to the physical state of the stored propellants. In this study, the design quality of interest ($Q_2$) for the propulsion sSE is chosen to be the delivered specific impulse ($I_{sp}$, measured in seconds) of solid propellants. Specific impulse is defined as the ratio of thrust to weight flow rate of the propellant and is a measure of energy content of the propellants~\cite{wertz_space_2011}. It signifies the energy to thrust conversion efficiency. $I_2$ is the cumulative investment on chemical propulsion research and technology by NASA over a period of ten fiscal years from 1979--1988.

\begin{figure}[h!]
\centering
\includegraphics[scale=1.0]{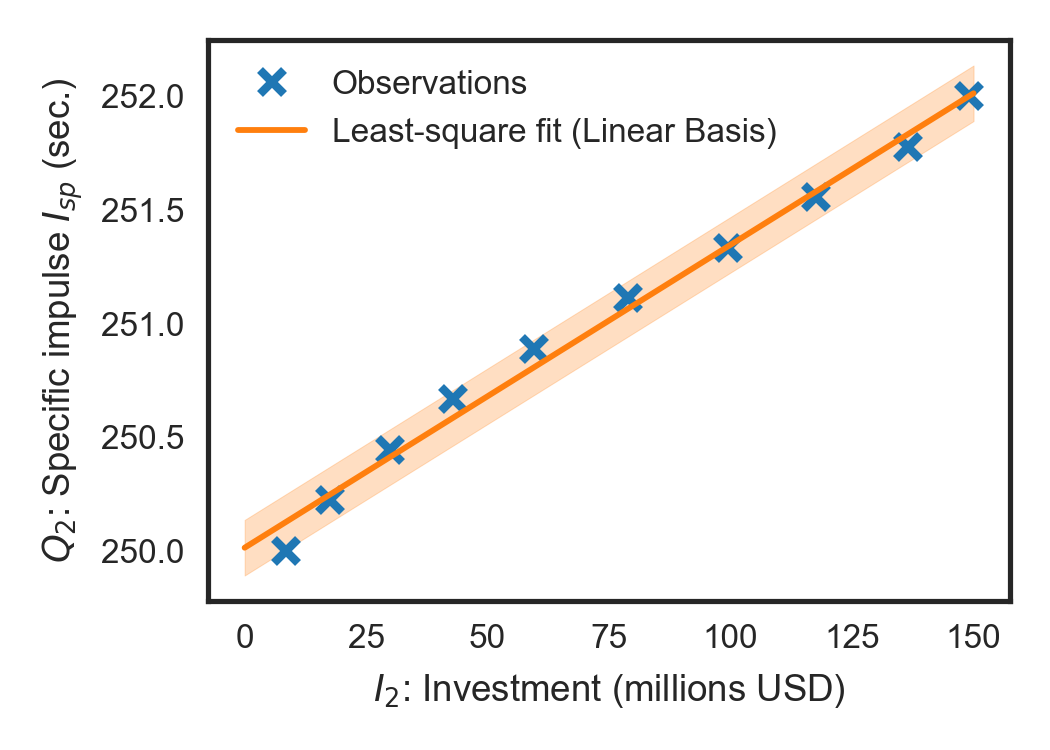}
 \caption{
Spacecraft case study (propulsion subsystem): Historical data (1979--1988) of specific impulse of solid mono-propellants vs cumulative investment per firm.
The solid line and the shaded area correspond to the maximum likelihood fit of a linear regression model and the corresponding $95\%$ prediction intervals, respectively. 
}
\label{propdata3}
\end{figure}

Trends in delivered specific impulse ($Q_2$ (sec.)) and investments by NASA ($I_2$ (millions USD)) in chemical propulsion technology with time are obtained from~\cite{noauthor_solid_nodate} and~\cite{noauthor_nasa_nodate}, respectively. The state-of-the-art solid propellant technology corresponds to a $Q_{20}$ value of 252~sec. and $I_{20}$ value of 149.1 million USD.
In this case, the maximum likelihood fit of the parameters results in a regression coefficient of $A_2=0.0133$ sec. per million USD, and standard deviation $\Sigma_2 = 0.12$ sec. The corresponding data and the maximum likelihood fit are illustrated in Fig.~\ref{propdata3}


The value of $C_2$ in this case is approximately same as that of $C_1$, i.e., $C_2 = $ 100,000 USD according to the data obtained from~\cite{prop_median_sal}. Substituting the values of $Q_{20}$, $C_2$ and $A_2$ in Eq.~\ref{ai_value} yields the following relationship between the parameters $Q_{2r}$, $T_2$ and $a_2$:
\begin{equation}
\label{a2}
a_{2} = \frac{0.0013 T_{2}}{Q_{2r}-252}  
\end{equation}

\begin{figure}[h!]
\centering
\includegraphics[scale=1.0]{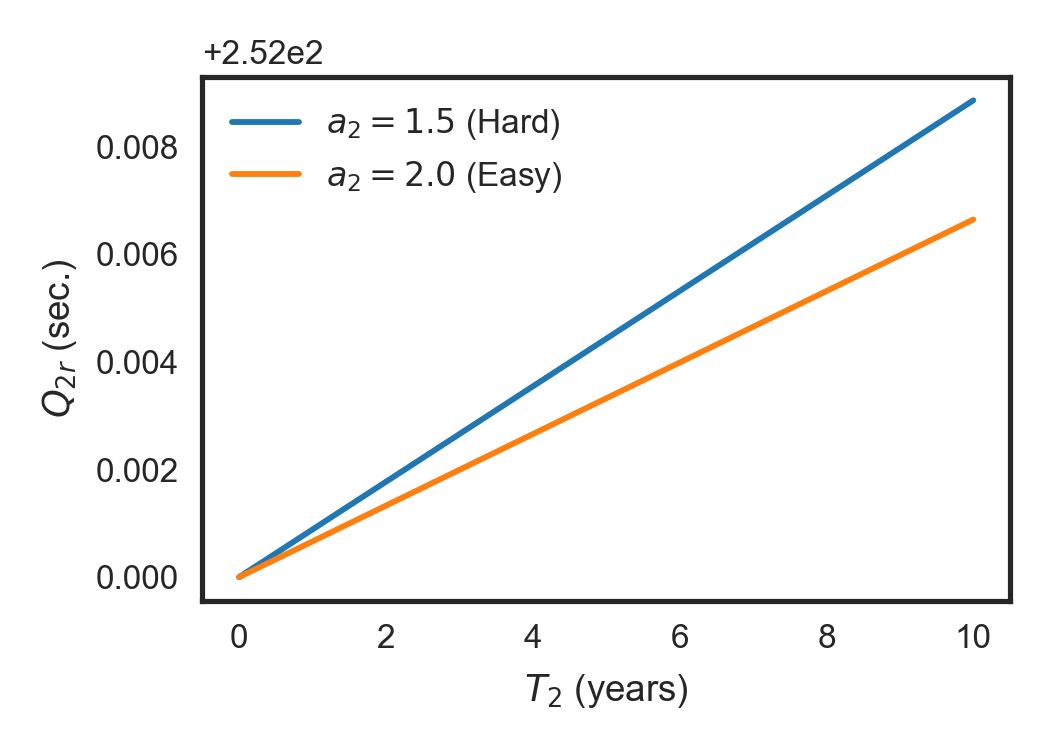}
 \caption{Spacecraft case study (propulsion subsystem): Variation of $Q_{2r}$ (Specific impulse) w.r.t $T_2$ (time in years)}
\label{propcontour}
\end{figure}

Fig.~\ref{propcontour} shows the variation of required specific impulse for sSE-~2 ($Q_{2r}$~(sec.)) with time for which one propulsion engineer has to be hired ($T_2$ years) for two different levels of subsystem design difficulty.

\section{Results}
\begin{figure*}
\centering
\begin{subfigure}{0.48\textwidth}
\centering
\includegraphics[width=\textwidth]{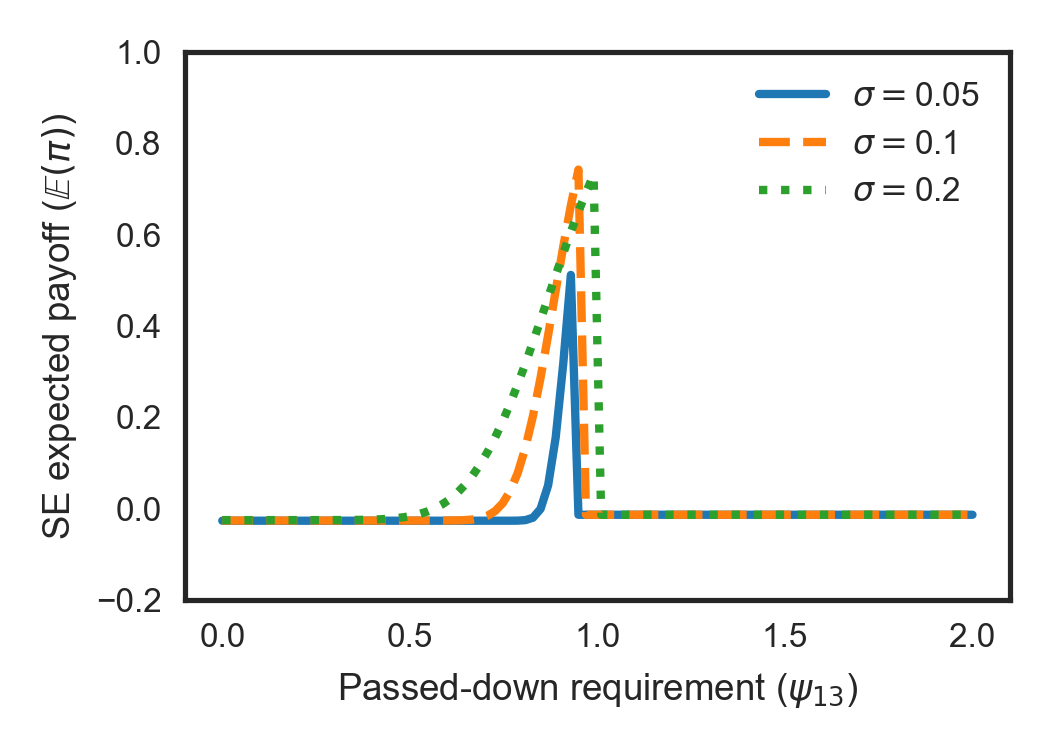}
\caption{Hard ($a_i=1.5$) - low cost ($c_i=0.01$).\\
$\sigma=0.05$: $\psi_1=\left(0.000, 0.007, \psi_{13}\right)$, $\psi_2=\left(0.000, 0.007, 0.945\right)$;\\
$\sigma=0.10$: $\psi_1=\left(0.001, 0.007, \psi_{13}\right)$, $\psi_2=\left(0.001, 0.007, 0.953\right)$;\\
$\sigma=0.20$: $\psi_1=\left(0.002, 0.007, \psi_{13}\right)$, $\psi_2=\left(0.002, 0.007, 0.968\right)$.}
\label{t1}
\end{subfigure}
\begin{subfigure}{0.48\textwidth}
\centering
\includegraphics[width=\textwidth]{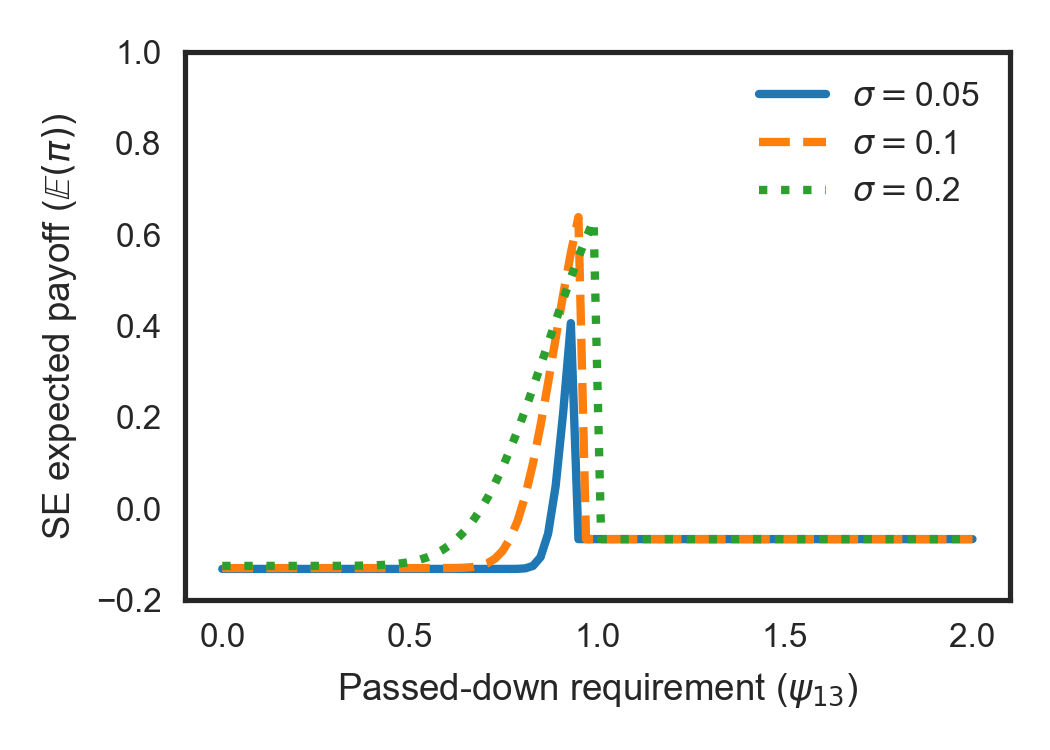}
\caption{Hard ($a_i=1.5$) - high cost ($c_i=0.05$).\\
$\sigma = 0.05$: $\psi_1=\left(0.004, 0.033, \psi_{13}\right)$, $\psi_2=\left(0.004, 0.033, 0.945\right)$;\\
$\sigma=0.10$: $\psi_1=\left(0.007, 0.033, \psi_{13}\right)$, $\psi_2=\left(0.007, 0.033, 0.953\right)$;\\
$\sigma=0.20$: $\psi_1=\left(0.012, 0.033, \psi_{13}\right)$, $\psi_2=\left(0.012, 0.033, 0.968\right)$.}
\label{t2}
\end{subfigure}
\begin{subfigure}{0.48\textwidth}
\centering
\includegraphics[width=\textwidth]{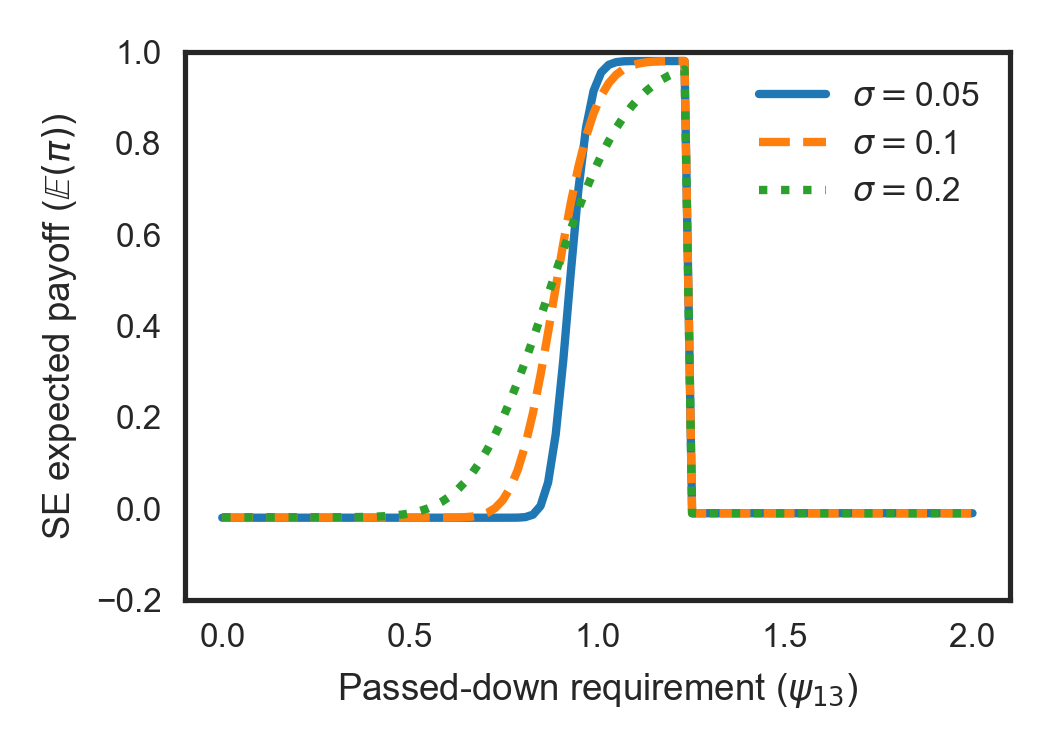}
\caption{Easy ($a_i=2.0$) - low cost ($c_i=0.01$).\\
$\sigma = 0.05$: $\psi_1=\left(0.001, 0.005, \psi_{13}\right)$, $\psi_2=\left(0.001, 0.005, 1.087\right)$;\\
$\sigma=0.10$: $\psi_1=\left(0.002, 0.005, \psi_{13}\right)$, $\psi_2=\left(0.002, 0.005, 1.205\right)$;\\
$\sigma=0.20$: $\psi_1=\left(0.003, 0.005, \psi_{13}\right)$, $\psi_2=\left(0.003, 0.005, 1.221\right)$.}
\label{t3}
\end{subfigure}
\quad
\begin{subfigure}{0.48\textwidth}
\centering
\includegraphics[width=\textwidth]{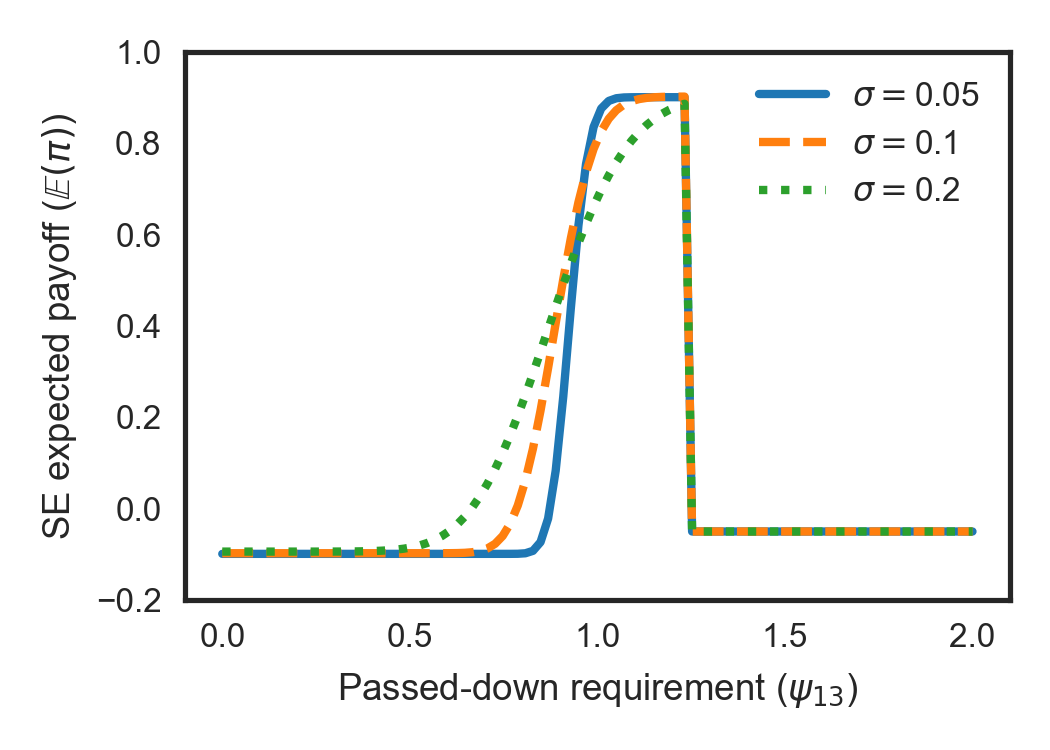}
\caption{Easy ($a_i=2.0$) - high cost ($c_i=0.05$).\\
$\sigma=0.05$: $\psi_1=\left(0.005, 0.025,\psi_{13}\right)$, $\psi_2=\left(0.005, 0.025, 1.087\right)$;
$\sigma=0.10$: $\psi_1=\left(0.009, 0.025, \psi_{13}\right)$, $\psi_2=\left(0.009, 0.025, 1.142\right)$;\\
$\sigma=0.20$: $\psi_1=\left(0.014, 0.025, \psi_{13}\right)$, $\psi_2=\left(0.014, 0.025, 1.221\right)$.}
\label{t4}
\end{subfigure}
\caption{The SE's expected payoff vs. passed-down requirement ($\psi_{13}$) for several types of sSE}
\label{pass}
\end{figure*}

We present numerical examples of one-shot SE processes with two identical sSEs ($N=2)$.
We investigate 4 different cases consisting of all possible combinations of two difficulty levels (easy ($a_i=2)$ vs hard ($a_i=1.5$)) and sSE opportunity cost levels (low ($c_i=0.01$) vs high ($c_i=0.05$)).
For each case, we consider three scenarios with different uncertainty levels: $\sigma = 0.05, 0.1$, and $0.2$.
For each scenario, we obtain the optimal contract $\boldsymbol{\psi}^*$ by solving the mechanism design problem of Sec.~II within the class of requirement-based incentives. 
Finally, we study the sensitivity of the SE's expected payoff on the passed-down requirement by changing the values of $\psi_{13}$ in the range $\left[0,2\right]$.
All numerical results can be found in Fig.~\ref{pass} and its captions.
Specifically, Figs.~\ref{t1}, \ref{t2}, \ref{t3}, and \ref{t4} include the results for a hard-task--low-cost-sSE, hard-task--high-cost-sSEs, easy-task--low-cost-sSE, and easy-task--high-cost-sSE, respectively.
The optimal contracts of each scenario can be read from the captions of the associated subfigures.

We start by commenting on the properties of the payment amounts $\psi_{i1}^*$ and $\psi_{i2}^*$.
We will refer to $\psi_{i1}^*$ as the participation payment, i.e., the fixed payment made to the sSE independently of the design outcome, and to $\psi_{i2}^*$ as the bonus payment, i.e., the payment made to the sSE if the passed-down requirements are met in our one-shot systems engineering model.
First, we observe that across all the scenarios the participation payment increases as uncertainty grows.
This makes sense, since the sSE is expected to ask for a higher certain gain to accept a higher risk task.
Counter-intuitively, the bonus payment is independent of the uncertainty level.
Second, for a fixed opportunity costs the participation payment decreases with increasing task difficulty while the bonus payment behaves in the opposite way (it increases with increasing task difficulty).
This means that for harder tasks the SE hedges themselves by shifting some of the participation payment to the bonus.
Finally, as the opportunity cost increases, both the participation and bonus payments increase.

Let us focus on the optimal passed-down requirements ($\psi_{i3}^*$).
Across all scenarios, the optimal passed-down requirement increases as the uncertainty becomes larger.
This is an example of the SE attempting to increase the probability of the actual requirement being met as uncertainty increases.
Moreover, we observe that the optimal passed-down requirement is independent of the opportunity cost.
This is due to the fact that the SE value function is requirement-based.
Furthermore, we observe that for easy tasks, captions of Figs.~\ref{t3} and \ref{t4}, the optimal passed-down requirement is always greater than the actual requirement $r_i=1$.
This result can be intuitively understood as follows.
Since the task is easy, the sSE will definitely reach the actual requirement with sufficient effort, even in the presence of significant uncertainties.
By setting the requirement threshold higher than the actual one, the SE forces the sSE to use more effort, effectively increasing the probability of success to almost certainty.
In contrast, for hard tasks, Figs~\ref{t1} and \ref{t2}, the optimal passed-down requirements are lower than the actual system requirement.
Since the task is hard, there is a high probability that the sSE may not be able to achieve a very high passed-down requirement.
To lure the sSEs to participate, the SE needs to lower the passed-down requirements below the actual threshold.
At a first glance, this may look like the SE will not be able to gain a positive expected payoff.
However, due to uncertainty in the final outcome, there is a significant probability that the sSE will produce better than the requested result.

Naturally, we observe that the expected SE payoff is increasing as the opportunity costs go down and that it decreases as the task difficulty increases.
For easy tasks, the expected SE payoff decreases as the uncertainty increases, Figs. \ref{t3} and \ref{t4}.
On the contrary, for hard tasks the expected SE payoff increases as a function of the uncertainty, Figs. \ref{t1} and \ref{t2}.
As we saw in the previous paragraph, this makes sense for a hard task that, an increase in uncertainty makes it more likely that the actual requirement is met. 

\section{Conclusion}
We presented a principle agent model of a one-shot, shallow systems engineering process.
We assumed that all agents are risk neutral and, thus, they maximize their expected payoff.
We modeled the quality function of subsystem engineers as a linear function of effort plus some Gaussian noise.
Using a spacecraft design case study, we demonstrated how the parameters of our model can be estimated from historical data.
Finally, we posed the optimal mechanism design problem within the class of requirement-based incentives.

Our one-shot model of systems engineering process challenges the intuitive belief that one should ask for higher requirements as the design task becomes more difficult.
Our model predicted that for a hard task, the optimal passed-down requirement should be less than the actual requirement.
The reason is that in this way the sSE is lured into participation while the SE may still meet the requirement because the design outcome may actually be better than anticipated.
Our result does not mean that this common belief is wrong.
After all, it is a very simple model, capturing only one iteration of systems engineering process.
This result may change if the quality function is not linear or if the design quality noise becomes skewed. 
Even if the modeling choices were spot on, at the present stage, it is too simple to be truly descriptive.
There may be mechanisms beyond the ones included in our model that incentivize the SE to ask for a requirement higher than what our model predicts?
One reason is that they may underestimate the difficulty of the task.
Another reason is that they may want to hedge themselves against dishonest behavior of sSEs, e.g., the sSE may put a design on their back pocket for later use.
Third, the systems engineering process may be taking place iteratively and asking for a higher requirement may be an effective way to probe what is possible.
From the perspective of the sSE, why would they accept to participate in a hard task with an exceedingly high requirement?
Of course, they may also underestimate the difficulty of the task.
Alternatively, they may be offered a participation payment that is high enough so that they do not care that it is impossible to meet the passed-down requirement.
Finally, they may believe that they will be able to renegotiate the contract in the future, especially if the SE has a history of doing so.
All these intricacies, and many more, are not captured by our model.
They are topic of on going research towards a theoretical foundation of systems engineering design that accounts for human behavior.

\section*{Acknowledgment}
This material is based upon work supported by the National Science Foundation under Grant No. 1728165.





\bibliographystyle{IEEEtran}
\bibliography{ref,ref_3}
%
%
%

\end{document}